\documentclass{ws-mpla}
\usepackage[super]{cite}
\usepackage{graphicx}

\usepackage{subfig}
\usepackage{physics}
\usepackage{bbold}
\usepackage[normalem]{ulem}
\usepackage{hyperref}
\usepackage{makecell}

\usepackage{tikz}
\usepackage[compat=1.1.0]{tikz-feynman}
\usetikzlibrary{external}
\begin{document}

\markboth{P. Dedin Neto, E. Kemp}{ Neutrino-(anti)neutrino forward scattering potential for massive neutrinos at low energies}

\catchline{}{}{}{}{}

\title{Neutrino-(anti)neutrino forward scattering potential for massive neutrinos at low energies}

\author{P. Dedin Neto}

\address{Gleb Wataghin Institute of Physics, University of Campinas, \\
R. Sérgio Buarque de Holanda, 777 ,Campinas, 13083-859, Brazil\\
dedin@ifi.unicamp.br}

\author{E. Kemp}

\address{Gleb Wataghin Institute of Physics, University of Campinas, \\
R. Sérgio Buarque de Holanda, 777 ,Campinas, 13083-859, Brazil\\
kemp@unicamp.br }

\maketitle

\pub{Received (Day Month Year)}{Revised (Day Month Year)}

\begin{abstract}
In this work, we calculate expression for the potential due to neutrino-(anti)neutrino forward scattering at low energies ($E<<m_{Z^0}$) for ultra-relativistic massive neutrinos ($E>>m_{\nu}$), a representative regime within astrophysical scenarios. There is a broadly used expression for this potential in the literature, which, however, lacks an explicit derivation from basic principles of quantum field theory. Therefore, this paper has the intention to guide the reader through the steps and concepts to derive this potential, trying to be clear and pedagogical. Moreover, we used a rigorous approach concerning the massive nature of the neutrinos, using massive quantized neutrino fields throughout the entire process, while the usual approach is to consider massless neutrino fields at the interaction. In this context, we explicitly show the validity of the massless neutrino fields approximation at the ultra-relativistic regime, as expected. As the last step, we connect the potential expression to the density matrix formalism, which is a usual framework for works considering neutrino-neutrino interactions. We also discuss some theoretical details through the paper, such as the normal ordering of quantum operators and the implications of massive fields in the neutrino state at its production.

\end{abstract}

\ccode{PACS Nos.: include PACS Nos.}

\section{Introduction}
\label{sec:intro}

\begin{figure}[hbt!]
 \centering
\subfloat[][Without flavor exchange.]{
\feynmandiagram [horizontal=i1 to i2] {
  i1 [particle=\(\nu_{e}(p_{1})\)] -- [fermion] a [dot] -- [fermion] i2 [particle=\(\nu_{\mu}(p_{2})\)],
  f1 [particle=\(\nu_{\mu}(p_{2})\)] -- [fermion] a -- [fermion] f2 [particle=\(\nu_{e}(p_{1})\)],
};
}
\subfloat[][With flavor exchange.]{
\feynmandiagram [horizontal=i1 to i2] {
  i1 [particle=\(\nu_{e}(p_{1})\)] -- [fermion] a [dot] -- [fermion] i2 [particle=\(\nu_{e}(p_{2})\)],
  f1 [particle=\(\nu_{\mu}(p_{2})\)] -- [fermion] a -- [fermion] f2 [particle=\(\nu_{\mu}(p_{1})\)],
};
}
\caption{Diagram of a neutrino-neutrino interaction ($\nu_e-\nu_\mu$) at low energies ($E_\nu<<m_{Z^0}$). At the left, we have a forward scattering without flavor exchange, and at the right, we have the equivalent with the flavor exchange.}
\label{fig:nu-nu_low_energy}
\end{figure}
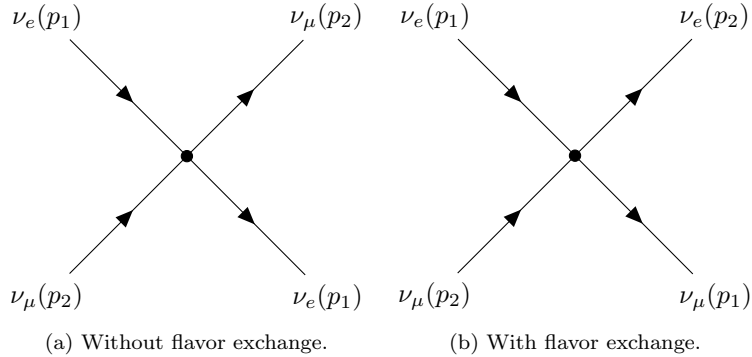

When neutrinos cross a matter profile, they are subject to a non-zero forward scattering potential due to charged leptons in the medium. That was first recognized by Wolfenstein \cite{wolfenstein1978neutrino},  who showed that this potential induces an effective mass in the neutrinos, affecting their oscillation. After that, Mikheyev and Smirnov \cite{mikheev1985resonance} showed that the matter profile along the neutrino trajectory could lead to resonant regions, where drastic modifications in the flavor transition probability occur. These works resulted in the well-established MSW effect. Some authors \cite{} noted that a similar effect could occur for neutrino-neutrino interactions while propagating in a medium with a high density of neutrinos, which can be the case for a supernova environment and the early universe as well. However, it was recognized later by Pantaleone \cite{pantaleone1992dirac,pantaleone1992neutrino} that the neutrino-neutrino interactions have off-diagonal elements in flavor basis (Figure \ref{fig:nu-nu_low_energy}), allowing the neutrino to exchange flavor in the interactions. Since then, a large collection of works have been published analyzing the effects of such neutrino-neutrino interactions. These effects are often called "collective oscillations " \cite{Duan:2010bg}, where the neutrinos evolve in a coupled manner. 

Although most of the papers in the field use an expression for the neutrino-neutrino effective potential, most of them do not show how this expression is derived, relying on the results of a limited number of sources. Among these sources, the main one is \cite{sigl1993general}, where the authors have an extensive discussion about the EoM (Equation of Motion) in a Heisenberg picture approach. However, the paper does not show explicitly how we can arrive at the potential matrix that rules the neutrino evolution (Equation 5.2 in \cite{sigl1993general}) starting from the interaction Hamiltonian (Equation 5.1 in \cite{sigl1993general}). Other works \cite{kostelecky1995self,fuller1987resonant} use an approach that averages the interaction Lagrangian or Hamiltonian over the background of neutrinos, similar to what is done for an electron background. However, it is not clear how this can be done, given that the neutrino fields in the Hamiltonian operator can act both in the state of the propagating neutrino or the state of background neutrinos. This average is allowed in the neutrino-electron interaction because the electron fields can act just on electron states. Other works \cite{n1988neutrono,d1994effect} derive the potential using Finite Temperature QFT (Quantum Field Theory), based on propagators in a thermal medium instead of the vacuum ones.

Given this picture, our intention in this paper is to derive the neutrino-neutrino potential at low energies (less than $Z^{0}$ mass) in the clearest way possible by showing the calculation steps and discussing the relevant concepts. We will work only on the ground of the standard QFT, trying to keep the document as accessible as possible. In this context, we will derive the potential using second quantized fields, in a similar way to what is done elsewhere \cite{giunti2007fundamentals, PDG_2020, Pasquini:2019tbe} for the electron-neutrino forward scattering potential. Moreover, we decided to be rigorous about the massive nature of the neutrino. To the best of the authors’ knowledge, this is the first time that these calculations are done explicitly using massive neutrino fields at the interaction. Previous works consider the neutrino field as massless at the interaction, which is a valid approximation in the ultra-relativistic regime, as we will show. In this way, we develop our calculations on a consistent ground regarding the massive nature of the neutrinos with no further complications.

We organize this paper as follows. In section \ref{sec:Nu-Nu Potential}, we calculate the forward scattering potential due to neutrino-neutrino interactions. In the first part of this section, we discuss the neutrino state at the production, its propagation, and the Hamiltonian governing its evolution. After this preliminary discussion, we calculate the neutrino-neutrino potential itself, followed by section \ref{sec:Neutrino-Antineutrino}, where we do the same calculations considering neutrino-antineutrino interactions. In the next section \ref{sec:Density_matrix}, we connect the potential expressions to the density matrix formalism, which is extensively used in the case of neutrino-neutrino phenomenology due to its connection to the polarization vector formalism. Finally, in section \ref{sec:Conclusions}, we draw our conclusions about this work.

\section{Neutrino-Neutrino Potential}
\label{sec:Nu-Nu Potential}

In this section, we derive the neutrino-neutrino forward scattering potential.  Before starting the calculations, we make a preliminary discussion about the neutrino state at the production,  its evolution, and possible interactions. Here we were strongly influenced by \cite{giunti2007fundamentals}, borrowing most of its notation.

\subsection{Preliminaries}

\subsubsection*{Neutrino state at the production}
The first problem that arises when working with massive neutrino fields is the definition of the neutrino state at the production. The reason is that the flavor fields that diagonalize the Standard Model interactions do not have well-defined masses. If we look at the production process of a neutrino  $\nu_\alpha$, like in (\ref{eq:Production_process}), its flavor is defined by the charged lepton partner $l^+_\alpha$ in the reaction\footnote{Neutrinos may also be produced by neutral current interactions. However, the flavor definition comes experimentally from charged current interactions, like the one in equation \ref{eq:Production_process}}.

\begin{equation}
\label{eq:Production_process}
P_I \rightarrow P_F + l^{+}_\alpha +\nu_\alpha
\end{equation}

where $P_I$ and $P_F$ are, respectively, the initial and final particles involved in the process. It is usual to consider the produced neutrino as a flavor eigenstate defined as follows.

\begin{equation}
\label{eq:flavor_state}
    \ket{\nu_\alpha} \equiv \sum_k U^*_{\alpha k} \ket{\nu_k}
\end{equation}

However, as mentioned before, since flavor fields do not have well-defined masses, the quantized flavor fields do not construct a well-behaved Fock space with physical masses \cite{Giunti:2003dg}, i.e., they cannot create or annihilate physical neutrino flavor states. The correct way is to consider the created neutrino $\nu_{\alpha}$ as a superposition of mass eigenstates $\ket{\nu_k}$, which come from quantized massive fields $\nu_k$ with well-defined masses. Each mass eigenstate have its own amplitude of production $\mathcal{A}^P_{\alpha k} = \bra{P_F + l^{+}_\alpha +\nu_k}\mathcal{S}\ket{P_I}$ given by the allowed interactions. Therefore, we can write the produced neutrino state as follows

\begin{equation}
    \ket{\nu^P_\alpha} = N \sum_{k} \mathcal{A}^P_{\alpha k} \ket{\nu_k} =  N \sum_{k} U^*_{\alpha k} {M}^P_{\alpha k} \ket{\nu_k}
\end{equation}

where $N$ is just a normalization constant and we have made the decomposition ${A}^P_{\alpha k} = U^*_{\alpha k} {M}^P_{\alpha k}$. Note that this state is generally not equal to the usual one defined in equation \ref{eq:flavor_state}. However, if the experiment measuring $\ket{\nu^P_\alpha (t)}$ is not sensible to the mass at the time of production, we can consider the production amplitude as the same for all mass eigenstates (${M}^P_{\alpha k} \approx {M}^P_{\alpha}$). Since this experimental limitation is always verified, we can safely use

\begin{equation}
    \ket{\nu^P_\alpha}  \approx  N {M}^P_{\alpha} \sum_{k} U^*_{\alpha k} \ket{\nu_k} = \sum_{k} U^*_{\alpha k} \ket{\nu_k}
\end{equation}

Therefore, the usual flavor eigenstate as in equation \ref{eq:flavor_state} consists in a good approximation \footnote{A deeper discussion in this topic can be found in \cite{giunti2007fundamentals} chapter 8.}. Being a bit more rigorous, as the neutrino will be a particle localized in space, a better description is given by considering the neutrino as a wave packet. This consideration reflects in an amplitude of production $\mathcal{A}^P_{\alpha k} (\vec{p},h)$ that depends on the uncertainties involved in the process.

\begin{equation}
    \ket{\nu^P_\alpha} = N \sum_{k,h} \int \frac{dp^3}{(2\pi)^3} \frac{1}{\sqrt{2E_k}} \mathcal{A}^P_{\alpha k} (\vec{p},h) \ket{\nu_k (E_k,\vec{p},h)}
\end{equation}

where $\vec{p}$ is the momentum, $E_k$ the energy, and $h$ the helicity. Again, if ${M}^P_{\alpha k} \approx {M}^P_{\alpha}$ we have the following common expression

\begin{equation}
\label{eq:wave-packet}
\begin{split}
    \ket{\nu^P_\alpha} &\approx N \sum_{h} \int \frac{dp^3}{(2\pi)^3} \frac{1}{\sqrt{2E_k}} \mathcal{M}^P_{\alpha} (\vec{p},h) \sum_{k} U^*_{\alpha k}\ket{\nu_k (E_k,\vec{p},h)}\\ &= \sum_{h} \int \frac{dp^3}{(2\pi)^3} \frac{1}{\sqrt{2E_k}} \varphi_{\alpha} (\vec{p},h) \ket{\nu_\alpha (\vec{p},h)}
\end{split}
\end{equation}

where $\varphi_{\alpha} (\vec{p},h)$ is the helicity dependent momentum distribution of the neutrino created with a charged lepton of flavor $\alpha$. As the reader can see, we arrived at the same expressions usually used for the produced neutrino state, considering it a plane wave or a wave packet. We just followed a more rigorous approach once we are interested in treating carefully the neutrino massive nature.

\subsubsection*{Neutrino Evolution}
In the Schrodinger picture, the neutrino evolution is described by the Hamiltonian

\begin{equation}
\label{eq:evolution}
    i\frac{d}{dt} \ket{\nu_\alpha^P (t)} =   H \ket{\nu_\alpha^P (t)}
\end{equation}

The Hamiltonian can be divided into a free and an interacting part. We can further split the last into a contribution due to interaction with ordinary matter (electrons, protons, and neutrons) and another due to interaction with the neutrinos in the environment.

\begin{equation}
    H = H_0 + H_{int} = H_{vac}+ V_{matter} + V_{\nu\nu}
\end{equation}

In this paper, we will focus on the neutrino-neutrino $V_{\nu\nu}$ contribution and derive its expression. Moreover, we will only consider the neutrino-neutrino forward scattering, which is equivalent to considering a free streaming neutrino propagating through a neutrino dense environment. 

At this point, one may be concerned about the spatial distribution of the neutrino state being evolved in the equation \ref{eq:evolution} and how this distribution may affect with which particles it is interacting. As we will see, the forward scattering potential is uniform over all the space. However, the neutrino will be limited to a finite space, given by the size of its wave packet, and will interact only with the particles in this range, i.e., when we have a wave packet overlap. As the neutrino propagates, the number of particles (neutrino or electrons) interacting with it may change, given an interacting Hamiltonian that depends on time over the neutrino trajectory.\footnote{For the reader interested in the microscopic behavior of the neutrino potential, in the ordinary matter case, we recommend the reference \cite{Akhmedov:2020vua}.}.

\subsubsection*{Neutrino Fields}
As we will discuss below, the Hamiltonians can be written in terms of the second quantized neutrino fields. Considering the neutrinos as Dirac particles with well-defined masses, the Fourier expansion is given by the following expressions \footnote{As we are in the Schrodinger picture, the operators are time-independent. Therefore, we can use the fields that are solutions of the free Dirac equation, given that we can make this picture equivalent to the Interaction picture at a given time $t_0$, which is described by free fields.}.

\begin{subequations}
\begin{equation}
\label{eq:field_expansion}
\begin{split}
        \nu_k(\vec{x})&= \sum_{\vec{p},s}  \frac{1}{\sqrt{2VE_{\vec{p}}}} \left [ a^s_k (\vec{p}) u^s_k (\vec{p}) e^{i\vec{p}\cdot\vec{x}}+  b^{s \dagger}_k ({\vec{p}}) v^s_k(\vec{p}) e^{-i\vec{p}\cdot\vec{x}}\right ]
\end{split}
\end{equation}

\begin{equation}
\label{eq:field_expansion_adj}
\begin{split}
        \overline{\nu}_k(\vec{x})&= \sum_{\vec{p},s}  \frac{1}{\sqrt{2VE_{\vec{p}}}} \left [ a^{s\dagger}_k (\vec{p}) \overline{u}^s_k (\vec{p}) e^{-i\vec{p}\cdot\vec{x}}+  b^{s }_k ({\vec{p}}) \overline{v}^s_k(\vec{p}) e^{i\vec{p}\cdot\vec{x}}\right ]
\end{split}
\end{equation}

\end{subequations}

In this paper, we adopt a normalization in a finite volume $V$ with periodic conditions in the Fourier expansion of the fields. Although it is a less realistic approach, it will give us properly normalized states which are better to work with. Moreover, our final results will be volume-independent, allowing us to take the infinite volume limit at the end. See appendix \ref{sec:Convetions} for more information about our conventions. 

The flavor fields are related to the massive fields in equation \ref{eq:field_expansion} and \ref{eq:field_expansion_adj} by the PMNS (\textit{Pontecorvo-Maki-Nakagawa-Sakata}) matrix which diagonalizes the standard model weak interactions.

\begin{subequations}
\begin{equation}
        \nu_\alpha (x) = \sum_{k} U_{\alpha k}^{*} \nu_k (x)
\end{equation}

\begin{equation}
        \overline{\nu}_\alpha (x) = \sum_{k} U_{\alpha k}  \overline{\nu}_k (x)
\end{equation}
\end{subequations}
As already mentioned, these flavor fields do not have well-defined mass and do not generate a physical Fock space of flavor states. One usual way to work around this issue is to consider massless neutrino fields at the interaction, so that the flavor fields have a well-defined mass ($m=0$). That is a good approximation, given that the mass contribution is suppressed by the neutrino energy in the ultra-relativistic limit, as we will see. However, we will use massive fields through the entire calculation process, taking the ultra-relativistic limit in the final expression.

\subsubsection*{Standard Model Interaction Hamiltonian}

To calculate the potential $V_{\nu\nu}$, we will use the effective Hamiltonian density $\mathcal{H}_{\nu\nu}(x)$ at low energies. This energy regime is representative of astrophysical scenarios, such as in a supernova. In the standard model, it is given by the following neutral current Hamiltonian density

\begin{equation}
\begin{split}
        \mathcal{H}_{eff}^{(NC)}(x) &= \sum_{\alpha,\beta} \frac{G_F}{\sqrt{2}} [\overline{\nu}_\alpha(x) \gamma^{\rho}(1-\gamma^5) \nu_\alpha(x)] [\overline{\nu}_\beta(x) \gamma_{\rho}(1-\gamma^5) \nu_\beta(x)] \\
        &=\frac{G_F}{\sqrt{2}} j^\rho_{Z,\nu} (x) j^{Z,\nu}_\rho(x)
\end{split}
\end{equation}

The neutrino neutral currents $j^\rho_{Z,\nu}(x)$ can be written in terms of the flavor fields $\nu_\alpha(x)$ or massive fields $\nu_k(x)$, using the PMNS matrix to rotate one into the other.

\begin{equation}
\begin{split}
        j^\rho_{Z,\nu} (x) &= \sum_\alpha \overline{\nu}_\alpha(x) \gamma^{\rho}(1-\gamma^5)\mathbf{\nu}_\alpha(x)\\
         &= \sum_{\alpha, k}  U_{\alpha k} U^{*}_{\alpha j} [\overline{\nu}_\alpha(x) \gamma^{\rho}(1-\gamma^5)\mathbf{\nu}_j(x)]\\
         &= \sum_{k}  \overline{\nu}_k(x) \gamma^{\rho}(1-\gamma^5)\mathbf{\nu}_k(x)\\
\end{split}
\end{equation}

That shows us that the neutral current is invariant under the rotation of the fields, as stated by the GIM mechanism \cite{Glashow:1970gm}. Therefore, we can write the interaction Hamiltonian in terms of the massive fields without changing its format.

\begin{equation}
\label{eq:Hamiltonian_mass}
\begin{split}
        \mathcal{H}_{eff}^{(NC)}(x) &= \sum_{i,j} \frac{G_F}{\sqrt{2}} [\overline{\nu}_i(x) \gamma^{\rho}(1-\gamma^5) \nu_i(x)] [\overline{\nu}_j(x) \gamma_{\rho}(1-\gamma^5) \nu_j(x)]\\
        &= \sum_{i,j} \frac{G_F}{\sqrt{2}} [\overline{\nu}_i(x) \Gamma^\rho \nu_i(x)] [\overline{\nu}_j(x) \Gamma_\rho \nu_j(x)]
\end{split}
\end{equation}

Usually, the notation $\Gamma^\rho$ represents any Dirac matrix multiplication. Here we use it just as a short notation, replacing it with the one corresponding to the SM weak neutral currents when necessary, as in the first line of equation \ref{eq:Hamiltonian_mass}. 

\subsubsection*{Normal Ordering}

When using the second quantized Hamiltonians described above, we may find divergent expectation values. That is a common problem in quantum field theory, and it comes from the ambiguity in the ordering of the operators when we are quantizing a classical theory\footnote{For example, the harmonic oscillator Hamiltonian, which inspires the quantization of fields, can be written as $H=(1/2) (\omega q - ip)(\omega q + ip)$ or $H=(1/2) (\omega q + ip)(\omega q - ip)$ because the variables commute in the classical theory. In quantum mechanics, one corresponds to $H=\hslash \omega a^{\dagger}a$ whereas the other to $H=\hslash \omega aa^{\dagger}$, which are incompatible by a factor of $\hslash \omega [a,a^{\dagger}]$. However, both are equal in the classical limit ($\hslash \rightarrow 0$) as expected. A good discussion on this can be found in section 2.3 of Professor David Tong  QFT notes \url{http://www.damtp.cam.ac.uk/user/tong/qft/qft.pdf}}. An example of this divergence occurs when one calculates the matter potential for antineutrinos as shown in equation \ref{eq:Normal_Ordering_Ex}.

\begin{equation}
\label{eq:Normal_Ordering_Ex}
    \bra{f} [\overline{\nu}(x) \Gamma^\rho \nu(x)] \ket{i} \propto \sum_{k,k'} \bra{f} b(k)b^{\dagger}(k')\ket{i} = \sum_{k,k'} \left(-\bra{f} b^{\dagger}(k')  b(k)\ket{i} + \bra{f} \delta_{kk'}\ket{i} \right)
\end{equation}

Due to the anti-commutation rule for the creation and annihilation operators, a term proportional to $\delta_{kk'}$ appears, which will be divergent and independent of the initial and final particles, i.e., it is no zero even for the vacuum. In the neutrino literature, it is usual to purely omit this divergent delta term without making any comment about it. However, there is a simple argument to ignore this infinity. What is physically meaningful is the energy above the vacuum, being irrelevant if we add the same constant, finite or not, to the energy of all the states. A usual prescription to get well-behaved operators in QFT, with zero vacuum expectation value, is to normal order them ($:A:$)\footnote{Operation where all the annihilation operators go to the right.}. In our case, the operator is the interaction Hamiltonian. Therefore, to avoid this problem with divergent vacuum expectation values, we will take the normal ordering of the Hamiltonian from the start of our calculations.

\begin{equation}
    \mathcal{H}_{\nu\nu}(x) \rightarrow :\mathcal{H}_{\nu\nu}(x):
\end{equation}

\subsection{The Neutrino-Neutrino potential}
\label{sec:Neutrino-Neutrino Matrix Element}

Completed our preliminary discussion aimed to establish a common ground in the topic, in this section we will calculate the potential due to neutrino-neutrino forward scattering. Although the neutrino is in general described by a wave packet due to uncertainties, as in equation \ref{eq:wave-packet}, the forward scattering implies no momentum exchange. Therefore, we can calculate the scattering potential for each momentum eigenstate individually. Considering a forward scattering $\nu_{\alpha} (p_1,s_1), \nu_{\alpha'} (p_2,s_2) \rightarrow \nu_{\beta} (p_1,s_1) , \nu_{\beta'} (p_2,s_2)$ of neutrinos with well-defined momenta $p_i$, helicity $h_i$ and flavor, we calculate the potential by the matrix element of the Hamiltonian given the initial and final states.

\begin{equation}
\label{eq:potential}
\begin{split}
    V_{\alpha \beta} &= \frac{1}{4}\sum_{s_1,s_2}\int dx  \frac{\bra{\nu_\beta (p_1,s_1) ,  \nu_{\beta'} (p_2,s_2)} :\mathcal{H}_{\nu\nu}(x):\ket{\nu_\alpha (p_1,s_1) , \nu_{\alpha'} (p_2,s_2)}}
    {\sqrt{\bra{i}\ket{i}\bra{f}\ket{f}}}\\
    &= \frac{1}{4}\sum_{s_1,s_2} \int dx \frac{G_F}{\sqrt{2}} \frac{\mathcal{V}_{\alpha \beta}(x)}
    {\sqrt{\bra{i}\ket{i}\bra{f}\ket{f}}}
\end{split}
\end{equation}

The non-primed indices $\alpha$ and  $\beta$ are respectively the initial and final flavors of the neutrino with momentum $p_1$, while the primed ones ($\alpha'$ and  $\beta'$) are related to the neutrino with momentum $p_2$. As usual, we have taken the average over the initial helicity state, given our ignorance about it. The inner product in the denominator is necessary for right normalization (more information in appendix \ref{sec:Potential Normalization}). However, as we will be using a finite volume normalization with periodic conditions, the normalization of the states will give a Kronecker delta and this denominator will simply be equal to one. 

Transforming the flavor states into mass eigenstates by using equation \ref{eq:flavor_state}, the potential density can be written as

\begin{equation}
\label{eq:nu-nu_potential_density_flavor}
\begin{split}
    &\mathcal{V}_{\alpha \beta}(x) =\\ &=\sum_{i'',j''} \bra{\nu_\beta (p_1,s_1) ,  \nu_{\beta'} (p_2,s_2)} :[\overline{\nu}_{i''}(x) \Gamma^{\rho} \nu_{i''}(x)] [\overline{\nu}_{j''}(x) \Gamma_{\rho} \nu_{j''}(x)]: \ket{\nu_\alpha (p_1,s_1) , \nu_{\alpha'} (p_2,s_2)}\\
    &=\sum_{i,j} \sum_{i',j'} U^*_{\alpha i}U^*_{\alpha' i'}U_{\beta j}U_{\beta' j'}\\
    &\times\sum_{i'',j''}\bra{\nu_j (p_1,s_1) ,  \nu_{j'} (p_2,s_2)}:[\overline{\nu}_{i''}(x) \Gamma^{\rho} \nu_{i''}(x)] [\overline{\nu}_{j''}(x) \Gamma_{\rho} \nu_{j''}(x)]: \ket{\nu_i (p_1,s_1) , \nu_{i'} (p_2,s_2)}\\
    &=\sum_{i,j} \sum_{i',j'} U^*_{\alpha i}U^*_{\alpha' i'}U_{\beta j}U_{\beta' j'} \mathcal{V}^{i',j'}_{i j}(x)\\
\end{split}
\end{equation}

If we focus on the term $\mathcal{V}^{i',j'}_{i j}(x)$ and make the Fourier expansion of the fields, only the terms that annihilate and create two neutrinos will survive ($\propto a^{\dagger}aa^{\dagger}a$).

\begin{equation}
\label{eq:density_potential_fourier}
\begin{split}
    &\mathcal{V}^{i',j'}_{i j}(x) =\\ &=\sum_{i'',j''}\bra{\nu_j (p_1,s_1) ,  \nu_{j'} (p_2,s_2)}:[\overline{\nu}_{i''}(x) \Gamma^{\rho} \nu_{i''}(x)] [\overline{\nu}_{j''}(x) \Gamma_{\rho} \nu_{j''}(x)]: \ket{\nu_i (p_1,s_1) , \nu_{i'} (p_2,s_2)}\\
    &= \sum_{i'',j''} \sum_{\substack{\vec{q} \vec{q'} \vec{k} \vec{k'}\\ s s' h h' }} \frac{e^{i x(q-q'+k-k')}}{\sqrt{2VE_{q}} \sqrt{2VE_{q'}} \sqrt{2VE_{k}} \sqrt{2VE_{k'}}} [\overline{u}_{i''}^{s}(q) \Gamma^\rho u_{i''}^{s'}(q')] [\overline{u}_{j''}^{h}(k) \Gamma_\rho u_{j''}^{h'}(k')]\\
    & \times \bra{\nu_j (p_1,s_1) ,  \nu_{j'} (p_2,s_2)} : a^{s \dagger}_{i''}(q) a^{s'}_{i''}(q) a^{h \dagger}_{j''}(k)a^{h'}_{j''}(k') : \ket{\nu_i (p_1,s_1) , \nu_{i'} (p_2,s_2)}\\
\end{split}
\end{equation}

Normal ordering the operators inside $::$, opening the definition of the states, and using anticommutation rules, we can write the matrix element as follows.

\begin{equation}
\begin{split}
&\bra{\nu_j (p_1,s_1) ,  \nu_{j'} (p_2,s_2)} : a^{s \dagger}_{i''}(q) a^{s'}_{i''}(q') a^{h \dagger}_{j''}(k)a^{h'}_{j''}(k') : \ket{\nu_i (p_1,s_1) , \nu_{i'} (p_2,s_2)}\\
=& - \bra{\nu_j (p_1,s_1) ,  \nu_{j'} (p_2,s_2)}  a^{s \dagger}_{i''}(q) a^{h \dagger}_{j''}(k) a^{s'}_{i''}(q') a^{h}_{j''}(k') \ket{\nu_{i'} (p_2,s_2) ,  \nu_{i} (p_1,s_1)}\\
= & - \bra{0} \uline{a^{s_1}_{j}(p_1) a^{s_2}_{j'}(p_2) a^{s \dagger}_{i''}(q) a^{h \dagger}_{j''}(k)} \dotuline{a^{s'}_{i''}(q') a^{h}_{j''}(k')  a^{s_2 \dagger}_{i'}(p_2) a^{s_1 \dagger}_{i}(p_1)}\ket{0}\\
= & - [( \uline{\delta_{i'' j'} \delta_{q p_2} \delta_{j'' j} \delta_{k p_1} - \delta_{i'' j} \delta_{q p_1} \delta_{j'' j'} \delta_{k p_2}}) (\dotuline{\delta_{j'' i'} \delta_{k' p_2} \delta_{i'' i} \delta_{q' p_1} - \delta_{j'' i} \delta_{k' p_1} \delta_{i'' i'} \delta_{q' p_2}} ) ]
\end{split}
\end{equation}

Where we have omitted the deltas in helicity for simplicity, given that it has the same behavior as the momentum index. We use the underline and dots just to help the reader identify from which commutation each delta came. Note that summing over the momentum (and helicity) of the fields, as requested in equation \ref{eq:density_potential_fourier}, there will be only the four terms listed in table \ref{tab:Neutrino-Neutrio_possibilities}. Two of them (1,4) correspond to an exchange of mass eigenstates between the interacting neutrinos  $\ket{\nu_i(p_1), \nu_{i'} (p_2)} \rightarrow \ket{\nu_{i'}(p_1), \nu_{i} (p_2)}$, where the neutrino with momentum $p_1$ change from $\ket{\nu_i}$ to $\ket{\nu_i'}$ \footnote{This can alternatively be seen as a momentum exchange, although we still in a forward scattering with a neutrino with momentum $p_1$ and another with momentum $p_2$}. The other two (2,3) correspond to no exchange of eigenstates $\ket{\nu_i(p1), \nu_{i'} (p_2)} \rightarrow \ket{\nu_i(p_1), \nu_{i'} (p_2)}$. The presence of two terms for each case came from the freedom of the neutrino field to act in $\ket{\nu_i(p_1)}$ or $\ket{\nu_{i'}(p_2)}$.

\begin{table}[ht]
\begin{center}
    \caption{Possibilities of annihilation and creation of the neutrinos in terms of the initial and final states.}
    \label{tab:Neutrino-Neutrio_possibilities}
    \begin{tabular}{|c|c|c|c|}
        \hline
         Numb.& $(\vec{q},\vec{k},\vec{q'},\vec{k'})$ &  $ \mathcal{V}^{i',j'}_{i j}(x)$& \makecell{Eigenstate\\ exchange}\\
        \hline
         1& $(\vec{p}_2,\vec{p}_1,\vec{p}_1,\vec{p}_2)$&  $ - [\overline{u}^{s_2}_{j'}(p_2) \Gamma^\rho u^{s_1}_{i}(p_1)] [\overline{u}^{s_1}_{j}(p_1) \Gamma_\rho u^{s_2}_{i'}(p_2)] \delta_{ji'}\delta_{j'i}$& yes\\
         \hline
         2& $(\vec{p}_2,\vec{p}_1,\vec{p}_2,\vec{p}_1)$& $+ [\overline{u}^{s_2}_{j'}(p_2) \Gamma^\rho u^{s_2}_{i'}(p_2)] [\overline{u}^{s_1}_{j}(p_1) \Gamma_\rho u^{s_1}_{i}(p_1)]  \delta_{ji}\delta_{j'i'}$& no\\
         \hline
         3& $(\vec{p}_1,\vec{p}_2,\vec{p}_1,\vec{p}_2)$& $ + [\overline{u}^{s_1}_{j}(p_1) \Gamma^\rho u^{s_1}_{i}(p_1)] [\overline{u}^{s_2}_{j'}(p_2) \Gamma_\rho u^{s_2}_{i'}(p_2)]\delta_{ji}\delta_{j'i'}$& no\\
         \hline
         4& $(\vec{p}_1,\vec{p}_2,\vec{p}_2,\vec{p}_1)$& $ - [\overline{u}^{s_1}_{j}(p_1) \Gamma^\rho u^{s_2}_{i'}(p_2)] [\overline{u}^{s_2}_{j}(p_2) \Gamma_\rho u^{s_1}_{i}(p_1)] \delta_{ji'}\delta_{j'i}$ & yes\\
         \hline
    \end{tabular}
\end{center}
\end{table}

As the reader can see, the eigenstate exchange terms (1,4) have different momenta spinors inside the same square bracket. However, to use the trace techniques (Appendix \ref{sec:trace_tec}), we have to have spinors with the same momenta in these square brackets. Therefore, we need to exchange spinors in the different square brackets. This operation will give us only a minus sign. The reader can see this by two reasoning: (i)When doing a Fierz transformation, we already account for a minus sign due to fermionic anti-commutation. Therefore, an equivalent transformation with spinors will have to have a minus sign ; (ii) If we do the Fierz transformation for these terms before expanding the fields, the order of the ladder operators would be three commutations from our ordering in equation \ref{eq:density_potential_fourier}, which differs from our results by a minus sign.

Going back to the potential density in equation \ref{eq:nu-nu_potential_density_flavor} and using the results above we have
\begin{equation}
\label{eq:nu-nu_dens_potential_spinors}
\begin{split}
   \mathcal{V}_{\alpha\beta}  (x)  &=\sum_{i,i'} \frac{U^*_{\alpha i}U_{\beta i}U^*_{\alpha' i'}U_{\beta' i'}}{2VE_{p_1} 2VE_{p2}} 2 [\overline{u}^{s_1}_{i}(p_1) \Gamma^\rho u^{s_1}_{i}(p_1)] [\overline{u}^{s_2}_{i'}(p_2) \Gamma_\rho u^{s_2}_{i'}(p_2)]\\
   &+\sum_{i,i'} \frac{U^*_{\alpha i}U_{\beta' i}U^*_{\alpha' i'}U_{\beta i'}}{2VE_{p_1} 2VE_{p2}} 2[\overline{u}^{s_1}_{i}(p_1) \Gamma^\rho u^{s_1}_{i}(p_1)] [\overline{u}^{s_2}_{i'}(p_2) \Gamma_\rho u^{s_2}_{i'}(p_2)]\\
\end{split}
\end{equation}

Note that the space dependence despairs in the potential. That happens because there is no momentum exchange in the forward scattering,  resulting in a potential that is uniform over space. For plane waves, the state of the particles spreads over all the space, and the interaction happens everywhere. However, for real particles, the interaction happens only when there is an overlap of their wave packets. When the overlap ceases, so does the interaction. Going back to equation \ref{eq:nu-nu_dens_potential_spinors}, we can use the trace techniques developed in the appendix \ref{sec:trace_tec} to simplify the spinor and Dirac matrices multiplications.

\begin{equation}
    \overline{u}^{(h)}(p) \gamma^{\mu}(1-\gamma^5)  u^{(h)}(p) = 2(p^{\mu} - m s^{\mu}_h)
\end{equation}

If we consider the ultra-relativistic limit $E>>m$, we may write

\begin{equation}
    2(p^{\mu} - m s^{\mu}_h)= 2\left (E-h|\vec{p}|,\vec{p}-h \frac{E\vec{p}}{|\vec{p}|}\right) \overset{E\approx|\vec{p}|}{\approx} \left\{\begin{matrix}
(4E,4\vec{p}),\;\;\;\;\;\text{ for }h=-1 \\ 
(2\frac{m^2}{E},0),\;\;\;\;\;\;\text{ for }h=+1 
\end{matrix}\right.
\end{equation}

As the neutrino masses are below the $eV$ scale \cite{Aghanim:2018eyx, Aker:2019uuj, Zyla:2020zbs}, we have $E>>m$ for the most relevant cases. Therefore, the positive helicity component is suppressed by a factor of $m^2/E$, as already expected for ultra-relativistic particles. Here we can see that the consideration of a massless neutrino field in the interaction used in some works \cite{sigl1993general, Volpe:2013uxl} is a good approximation in this energy regime. Therefore, we can write the potential in equation \ref{eq:potential} as follows.

\begin{equation}
\begin{split}
    V_{\alpha \beta} &=  \sqrt{2} G_F\int dx   \sum_{i,i'} [U^*_{\alpha i}U_{\beta i}U^*_{\alpha' i'}U_{\beta' i'}+U^*_{\alpha i}U_{\beta' i}U^*_{\alpha' i'}U_{\beta i'}]\frac{(p_1^\rho - m_i s^{\rho}_{s_1})(p_{2\rho} - m_j s_{\rho}^{s_2})}{2VE_{p_1} 2VE_{p2}}\\
    &\overset{E\approx|\vec{p}|}{\approx}  \sqrt{2} G_F\int dx   \sum_{i,i'} [U^*_{\alpha i}U_{\beta i}U^*_{\alpha' i'}U_{\beta' i'}+U^*_{\alpha i}U_{\beta' i}U^*_{\alpha' i'}U_{\beta i'}]\frac{4(E_{p_1}E_{p_2} - \vec{p}_1 \cdot \vec{p}_2)}{2VE_{p_1} 2VE_{p2}}\\
    &=\frac{\sqrt{2}G_F}{V}(\delta_{\beta \alpha}\delta_{\beta' \alpha'} + \delta_{\beta \alpha'} \delta_{\beta' \alpha}) (1-\hat{p}_1 \cdot \hat{p}_2)
\end{split}
\end{equation}

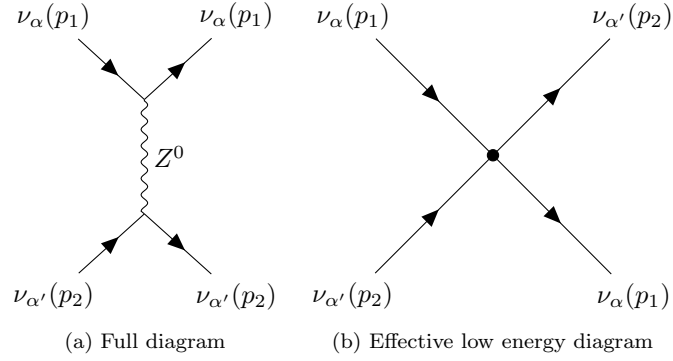
\begin{figure}[hbt!]
 \centering
 \subfloat[][Full diagram]{
\feynmandiagram [vertical=a to b] {
  i1 [particle=\(\nu_{\alpha}(p_{1})\)] -- [fermion] a -- [fermion] f1 [particle=\(\nu_{\alpha}(p_{1})\)],
  a -- [boson, edge label=\(Z^{0}\)] b,
  b -- [fermion] i2 [particle=\(\nu_{\alpha'}(p_{2})\)],
  f2 [particle=\(\nu_{\alpha'}(p_{2})\)] -- [fermion] b,
};
}
\subfloat[][Effective low energy diagram]{
 \feynmandiagram [horizontal=i1 to i2] {
  i1 [particle=\(\nu_{\alpha}(p_{1})\)] -- [fermion] a [dot] -- [fermion] i2 [particle=\(\nu_{\alpha'}(p_{2})\)],
  f1 [particle=\(\nu_{\alpha'}(p_{2})\)] -- [fermion] a -- [fermion] f2 [particle=\(\nu_{\alpha}(p_{1})\)],
};
}
\caption{Diagrams that represent the neutrino-neutrino forward scattering where there is no exchange of flavor between $\nu_\alpha (p_1)$ and $\nu_{\alpha'}(p_2)$.}
\label{fig:nu-nu_diagonal}
\end{figure}

\begin{figure}[hbt!]
 \centering
\subfloat[][Full diagram]{
\begin{tikzpicture}
\begin{feynman}
\diagram [vertical=a to b] {
i1 [particle=\(\nu_{\alpha}(p_{1})\)]
-- [fermion] a
-- [draw=none] f1 [particle=\(\nu_{\alpha'}(p_{1})\)],
a -- [photon, edge label'=\(Z^{0}\)] b,

b -- [draw=none] f2 [particle=\(\nu_{\alpha}(p_{2})\)],

i2 [particle=\(\nu_{\alpha'}(p_{2})\)] -- [fermion] b,
};

\diagram* {
(a) -- [fermion] (f2),
(b) -- [fermion] (f1),
};

\end{feynman}
\end{tikzpicture}
}
 \subfloat[][Effective low energy diagram]{
\feynmandiagram [horizontal=i1 to i2] {
  i1 [particle=\(\nu_{\alpha}(p_{1})\)] -- [fermion] a [dot] -- [fermion] i2 [particle=\(\nu_{\alpha}(p_{2})\)],
  f1 [particle=\(\nu_{\alpha'}(p_{2})\)] -- [fermion] a -- [fermion] f2 [particle=\(\nu_{\alpha'}(p_{1})\)],
};
}
\caption{Diagrams that represent the neutrino-neutrino forward scattering where there is an exchange of flavor between $\nu_\alpha(p_1)$ and $\nu_{\alpha'}(p_2)$.}
\label{fig:nu-nu_off_diagonal}
\end{figure}
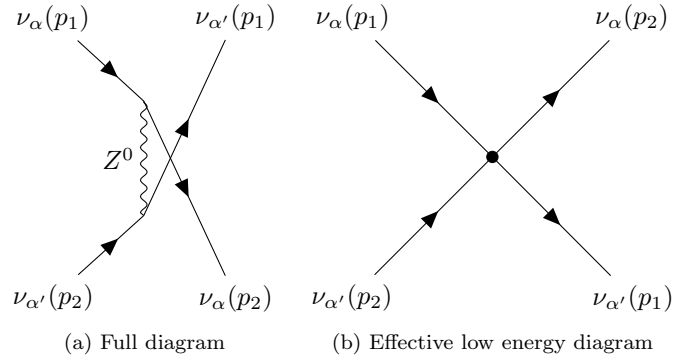

Where we have used the unitarity of the PMNS matrix $UU^{\dagger} = \mathbb{1}$. We can see, as pointed by J. Pantaleone \cite{pantaleone1992dirac,pantaleone1992neutrino}, that in the flavor basis the potential have diagonal $\delta_{\beta \alpha}$ and off-diagonal  $\delta_{\beta \alpha'}$ elements (Figures \ref{fig:nu-nu_diagonal} and \ref{fig:nu-nu_off_diagonal} respectively). As already found by \cite{n1988neutrono}, the potential felt by a neutrino interacting with a neutrino with the same flavor ($\alpha = \alpha'$) is twice the one felt in the interaction with another flavor ($\alpha \neq \alpha'$). That happens because the process that exchanges the flavor is indistinguishable from the one where there is no exchange, given that the interacting neutrinos have the same flavor.

Considering a propagating neutrino $\nu_\alpha (p_1,s_1)$, we can write the potential as a matrix in the flavor space of this neutrino, which will depend on the other neutrino flavor $\alpha'$.

\begin{equation}
\label{eq:Final_potential_nu_nu_with_diagonal}
V_{\nu\nu} = \frac{ \sqrt{2} G_F}{V} (1-\hat{p}_1 \cdot \hat{p}_2)
\begin{pmatrix}
 2\delta_{\alpha',e} + \delta_{\alpha',\mu} + \delta_{\alpha',\tau} & \delta_{\alpha',e} &\delta_{\alpha',e} \\ 
\delta_{\alpha',\mu} &  2 \delta_{\alpha',\mu} + \delta_{\alpha',e} + \delta_{\alpha',\tau} & \delta_{\alpha',\mu}\\ 
\delta_{\alpha',\tau} &\delta_{\alpha',\tau}  & 2\delta_{\alpha',\tau} + \delta_{\alpha',e} + \delta_{\alpha',\mu}
\end{pmatrix}
\end{equation}

Note that we may remove a diagonal term $\mathbb{1}_{3\times3}(\delta_{\alpha',e} + \delta_{\alpha',\mu} + \delta_{\alpha',\tau})$ that does not contribute to the phase difference in the evolution, but only to an overall phase.

\begin{equation}
\label{eq:Final_potential_nu_nu}
V'_{\nu\nu} = \frac{\sqrt{2}G_F}{V} (1-\hat{p}_1 \cdot \hat{p}_2)
\begin{pmatrix}
 \delta_{\alpha',e}& \delta_{\alpha',e} & \delta_{\alpha',e}\\ 
 \delta_{\alpha',\mu}& \delta_{\alpha',\mu} & \delta_{\alpha',\mu}\\ 
 \delta_{\alpha',\tau}& \delta_{\alpha',\tau} & \delta_{\alpha',\tau}
\end{pmatrix}
\end{equation}

Although this expression for the potential is similar to the one shown by J. Pantaleone \cite{pantaleone1992dirac,pantaleone1992neutrino}, the recent literature usually uses the one in terms of the density matrix. To find this common expression, we have to connect the potential to the density matrix formalism, what we do in section \ref{sec:Density_matrix}.  Before doing this, we will do the same calculations for the neutrino-antineutrino interaction, which has its peculiarities.

\section{Neutrino-Antineutrino Potential}
\label{sec:Neutrino-Antineutrino}
In the case of a neutrino $\nu_\alpha (p_1)$ interacting with an antineutrinos $\overline{\nu}_{\alpha'} (p_2)$, we only need to change the initial and final states in our potential density expression.

\begin{equation}
\label{eq:nu-antinu_potential_density_flavor}
\begin{split}
    &\mathcal{V}_{\alpha \beta}(x) =\\ &=\sum_{i'',j''} \bra{\nu_\beta (p_1,s_1) ,  \overline{\nu}_{\beta'} (p_2,s_2)} :[\overline{\nu}_{i''}(x) \Gamma^{\rho} \nu_{i''}(x)] [\overline{\nu}_{j''}(x) \Gamma_{\rho} \nu_{j''}(x)]: \ket{\overline{\nu}_{\alpha'} (p_2,s_2), \nu_\alpha (p_1,s_1)}\\
    &=\sum_{i,j} \sum_{i',j'} U^*_{\alpha i}U_{\alpha' i'}U_{\beta j}U^*_{\beta' j'}\\
    &\times\sum_{i'',j''}\bra{\nu_j (p_1,s_1) ,  \overline{\nu}_{j'} (p_2,s_2)}:[\overline{\nu}_{i''}(x) \Gamma^{\rho} \nu_{i''}(x)] [\overline{\nu}_{j''}(x) \Gamma_{\rho} \nu_{j''}(x)]: \ket{ \overline{\nu}_{i'} (p_2,s_2), \nu_i (p_1,s_1)}\\
    &=\sum_{i,j} U^*_{\alpha i}U_{\alpha' i'}U_{\beta j}U^*_{\beta' j'} \mathcal{V}^{i',j'}_{i j}(x)\\
\end{split}
\end{equation}
Focusing on the term $\mathcal{V}^{i',j'}_{i j}(x)$ and making the Fourier expansion of the fields, only the terms that annihilate/create a neutrino and an antineutrino will survive ($\propto a^{\dagger}ab^{\dagger}b$). That will make the calculations a bit different from the neutrino-neutrino case because the neutrino field can either annihilate a neutrino or create an antineutrino. The opposite occurs for its adjoint. With that in mind, the calculations go as follows.
\begin{equation}
\begin{split}
    &\mathcal{V}^{i',j'}_{i j}(x) =\\ &=\sum_{i'',j''}\bra{\nu_j (p_1,s_1) ,  \overline{\nu}_{j'} (p_2,s_2)}:[\overline{\nu}_{i''}(x) \Gamma^{\rho} \nu_{i''}(x)] [\overline{\nu}_{j''}(x) \Gamma_{\rho} \nu_{j''}(x)]: \ket{\overline{\nu}_{i'} (p_2,s_2), \nu_i (p_1,s_1)}\\
    &= \sum_{i'',j''} \sum_{\substack{\vec{q} \vec{q'} \vec{k} \vec{k'}\\ s s' h h' }} \frac{e^{i x(q-q'+k-k')}}{\sqrt{2VE_{q}} \sqrt{2VE_{q'}} \sqrt{2VE_{k}} \sqrt{2VE_{k'}}} \left \{ [\overline{u}_{i''}^{s}(q) \Gamma^\rho u_{i''}^{s'}(q')] [\overline{v}_{j''}^{h}(k) \Gamma_\rho v_{j''}^{h'}(k')] \mathcal{A}_{q q' k k'}^{i''j''} \right.\\
    &+ \left.
    [\overline{u}_{i''}^{s}(q) \Gamma^\rho u_{j''}^{s'}(k')][\overline{v}_{j''}^{h}(k) \Gamma_\rho v_{i''}^{h'}(q')] \mathcal{A}_{q q' k k'}^{i''j''} \right\}
\end{split}
\end{equation}
Where we already consider a Fierz transformation ($\nu_{i''}(q')\leftrightharpoons \nu_{j''}(k')$) when necessary to separate the particles and antiparticles spinors inside different square brackets. That will be useful to apply the trace techniques of appendix \ref{sec:trace_tec}. Note that in the last line we separate the terms with (right) and without (left) a Fierz transformation. As in the neutrino-neutrino case, there are four possible terms, according to the choice of which state a field will act on. We list them in the table \ref{tab:Antinu_contractions}, where the possibilities (3,4) are the ones with Fierz transformation.

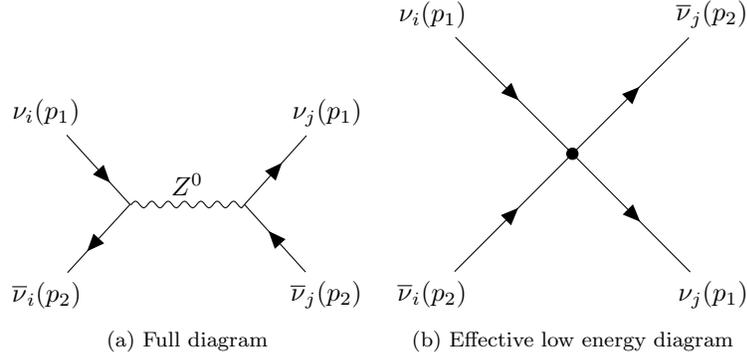
\begin{figure}[hbt!]
 \centering
\subfloat[][Full diagram]{
\feynmandiagram [horizontal=a to b]{
  i1 [particle=\(\overline{\nu}_{i}(p_{2})\)] -- [anti fermion] a -- [anti fermion] f1 [particle=\(\nu_{i}(p_{1})\)],
  
  a -- [boson, edge label=\(Z^{0}\)] b,
  b -- [fermion] i2 [particle=\(\nu_{j}(p_{1})\)],
  
  f2 [particle=\(\overline{\nu}_{j}(p_{2})\)] -- [fermion] b,
};
}
 \subfloat[][Effective low energy diagram]{
\feynmandiagram [horizontal=i1 to i2] {
  i1 [particle=\(\nu_{i}(p_{1})\)] -- [fermion] a [dot] -- [fermion] i2 [particle=\(\overline{\nu}_{j}(p_{2})\)],
  f1 [particle=\(\overline{\nu}_{i}(p_{2})\)] -- [fermion] a -- [fermion] f2 [particle=\(\nu_{j}(p_{1})\)],
};
}
\caption{Annihalation of a neutrino and an antineutrino with the same mass $m_i$ which allows the creation of a pair $\nu-\overline{\nu}$ of any mass $m_j$.}
\label{fig:nu-antinu_new_mass}
\end{figure}

 \begin{table}[ht]
\begin{center}
    \caption{Possibilities of annihilation and creation of the neutrino and antineutrino in terms of the initial and final states.}
    \begin{tabular}{|c|c|c|}
        \hline
         Numb.& $(\vec{q},\vec{q'},\vec{k},\vec{k'})$ &  $ \mathcal{A}_{q q' k k'}^{i''j''}$ \\
        \hline
        1& $(\vec{p}_1,\vec{p}_1,\vec{p}_2,\vec{p}_2)$ &  \makecell{$\bra{f} : a^{s \dagger}_{i''}(q) a^{s'}_{i''}(q') b^{h}_{j''}(k)b^{h'\dagger}_{j''}(k') : \ket{i}$=\\$-(\delta_{ji} \delta_{j'i'})(\delta_{q p_1}\delta_{q' p_1}\delta_{k p_2}\delta_{k' p_2})$}\\
        \hline
        2&$(\vec{p}_2,\vec{p}_2,\vec{p}_1,\vec{p}_1)$ & \makecell{$\bra{f} : b^{s }_{i''}(q) b^{s'\dagger }_{j''}(q') a^{h\dagger}_{j''}(k)  a^{h'}_{i''}(k') :  \ket{i} $=\\ $-(\delta_{ji} \delta_{ji'})(\delta_{q p_2}\delta_{q' p_2}\delta_{k p_1}\delta_{k' p_1})$}\\
        \hline
        3&$(\vec{p}_1,\vec{p}_2,\vec{p}_2,\vec{p}_1)$  &\makecell{$\bra{f} : a^{s \dagger}_{i''}(q) a^{h'}_{j''}(k') b^{h}_{j''}(k)b^{h'\dagger}_{i''}(q') : \ket{i}$=\\$-(\delta_{ii'} \delta_{jj'})(\delta_{q p_1}\delta_{q' p_2}\delta_{k p_2}\delta_{k' p_1})$}\\
        \hline
        4&$(\vec{p}_2,\vec{p}_1,\vec{p}_1,\vec{p}_2)$  &\makecell{$\bra{f} : b^{s}_{i''}(q) b^{h'\dagger }_{j''}(k') a^{h\dagger}_{j''}(k)  a^{s'}_{i''}(q') :  \ket{i} $= \\$-(\delta_{ii'} \delta_{jj'})(\delta_{q p_2}\delta_{q' p_1}\delta_{k p_1}\delta_{k' p_2})$}\\
         \hline
    \end{tabular}
    \label{tab:Antinu_contractions}
\end{center}
\end{table}
Going back to the potential density in equation \ref{eq:nu-antinu_potential_density_flavor} and using the results above, we have the following expression. 
\begin{equation}
\label{eq:V_density_antinu}
\begin{split}
   \mathcal{V}_{\alpha\beta}  (x)  =&-\sum_{i,i'} \frac{U^*_{\alpha i}U_{\alpha' i'}U_{\beta i}U^*_{\beta' i'}}{2VE_{p_1} 2VE_{p2}} 2 [\overline{u}^{s_1}_{i}(p_1) \Gamma^\rho u^{s_1}_{i}(p_1)] [\overline{v}^{s_2}_{i'}(p_2) \Gamma_\rho v^{s_2}_{i'}(p_2)]\\
   &-\sum_{i,j} \frac{U^*_{\alpha i}U_{\alpha' i}U_{\beta j}U^*_{\beta' j}}{2VE_{p_1} 2VE_{p2}} 2[\overline{u}^{s_1}_{i}(p_1) \Gamma^\rho u^{s_1}_{j}(p_1)] [\overline{v}^{s_2}_{j}(p_2) \Gamma_\rho v^{s_2}_{i}(p_2)]\\
\end{split}
\end{equation}
One difficulty that arises in the equation \ref{eq:V_density_antinu} is the presence of (anti)particle spinors with different masses in the last line, forbidding us to use the properties developed in appendix \ref{sec:trace_tec}. These terms come from the process where we annihilate a particle-antiparticle pair of equal mass $i=i'$ and create another pair of any mass $j=j'$, as indicated in figure \ref{fig:nu-antinu_new_mass}. 
One way to get around this problem is to consider that the masses are about the same $m_i\approx m_j \approx m_{\overline{\nu}}$. In any case, its contribution will be suppressed by the energy as already found in the neutrino-neutrino potential. Making this consideration, we can use the following relations from appendix \ref{sec:trace_tec}.
\begin{equation}
\label{eq:anti-spinor_trace_tec}
    \overline{v}^{(h)}(p) \gamma^{\mu}(1-\gamma^5)  v^{(h)}(p) = 2(p^{\mu} + m_{\overline{\nu}} s^{\mu}_h)
\end{equation}
\begin{equation}
    2(p^{\mu} + m_{\overline{\nu}} s^{\mu}_h)= 2\left (E+h|\vec{p}|,\vec{p}+h \frac{E\vec{p}}{|\vec{p}|}\right) \overset{E\approx|\vec{p}|}{\approx} \left\{\begin{matrix}
(4E,4\vec{p}),\;\;\;\;\;\text{ for }h=+1 \\ 
(2\frac{m_{\overline{\nu}}^2}{E},0),\;\;\;\;\;\;\text{ for }h=-1 
\end{matrix}\right.
\end{equation}
Considering the neutrinos as ultra-relativistic, the potential in equation \ref{eq:potential}, for the case of neutrino-antineutrino interactions, can be written as
\begin{equation}
\begin{split}
    V_{\alpha \beta} &= - \sqrt{2} G_F\int dx   \sum_{i,i'} [U^*_{\alpha i}U_{\alpha' i'}U_{\beta i}U^*_{\beta' i'}+U^*_{\alpha i}U_{\alpha' i}U_{\beta i'}U^*_{\beta' j}] \frac{(p_1^\rho - m_\nu s^{\rho}_{s_1})(p_{2\rho} + m_{\overline{\nu}} s_{\rho}^{s_2})}{2VE_{p_1} 2VE_{p2}}\\
    &\overset{E\approx|\vec{p}|}{\approx} - \sqrt{2} G_F\int dx   \sum_{i,i'} [U^*_{\alpha i}U_{\alpha' i'}U_{\beta i}U^*_{\beta' i'}+U^*_{\alpha i}U_{\alpha' i}U_{\beta i'}U^*_{\beta' j}] \frac{4(E_{p_1}E_{p_2} - \vec{p}_1 \cdot \vec{p}_2)}{2VE_{p_1} 2VE_{p2}}\\
    &=-\frac{\sqrt{2}G_F}{V}(\delta_{\beta \alpha} \delta_{\beta' \alpha'} + \delta_{\alpha \alpha'} \delta_{\beta \beta'}) (1-\hat{p}_1 \cdot \hat{p}_2)
\end{split}
\end{equation}

\begin{figure}[hbt!]
 \centering
 \subfloat[][Full diagram]{
\feynmandiagram [vertical=a to b] {
  i1 [particle=\(\nu_{\alpha}(p_{1})\)] -- [fermion] a -- [fermion] f1 [particle=\(\nu_{\alpha}(p_{1})\)],
  a -- [boson, edge label=\(Z^{0}\)] b,
  b -- [anti fermion] i2 [particle=\(\overline{\nu}_{\alpha'}(p_{2})\)],
  f2 [particle=\(\overline{\nu}_{\alpha'}(p_{2})\)] -- [anti fermion] b,
};
}
\subfloat[][Effective low energy diagram]{
 \feynmandiagram [horizontal=i1 to i2] {
  i1 [particle=\(\nu_{\alpha}(p_{1})\)] -- [fermion] a [dot] -- [fermion] i2 [particle=\(\nu_{\alpha}(p_{1})\)],
  f1 [particle=\(\overline{\nu}_{\alpha'}(p_{2})\)] -- [fermion] a -- [fermion] f2 [particle=\(\overline{\nu}_{\alpha'}(p_{2})\)],
};
}
\caption{Diagrams that represent the neutrino-antineutrino interaction where there is no exchange of flavor between $\nu_\alpha(p_1)$ and $\overline{\nu}_{\alpha'}(p_2)$}
\label{fig:nu-anti_nu_same_flavor}
\end{figure}
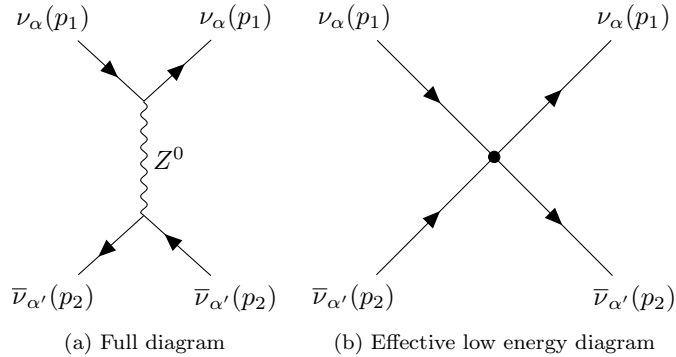

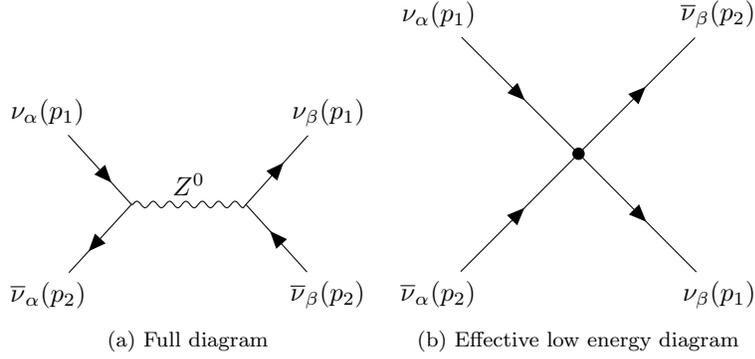
\begin{figure}[hbt!]
 \centering
\subfloat[][Full diagram]{
\feynmandiagram [horizontal=a to b]{
  i1 [particle=\(\overline{\nu}_{\alpha}(p_{2})\)] -- [anti fermion] a -- [anti fermion] f1 [particle=\(\nu_{\alpha}(p_{1})\)],
  
  a -- [boson, edge label=\(Z^{0}\)] b,
  b -- [fermion] i2 [particle=\(\nu_{\beta}(p_{1})\)],
  
  f2 [particle=\(\overline{\nu}_{\beta}(p_{2})\)] -- [fermion] b,
};
}
 \subfloat[][Effective low energy diagram]{
\feynmandiagram [horizontal=i1 to i2] {
  i1 [particle=\(\nu_{\alpha}(p_{1})\)] -- [fermion] a [dot] -- [fermion] i2 [particle=\(\overline{\nu}_{\beta}(p_{2})\)],
  f1 [particle=\(\overline{\nu}_{\alpha}(p_{2})\)] -- [fermion] a -- [fermion] f2 [particle=\(\nu_{\beta}(p_{1})\)],
};
}
\caption{Neutrino-antineutrino annihilation which allows the creation of a new flavor $\beta$, which may be equal to the initial one $\beta=\alpha$ or different $\beta\neq\alpha$.}
\label{fig:nu-anti_nu_new_flavor}
\end{figure}

Therefore, there are a flavor diagonal contribution $\delta_{\beta\alpha}$ (Figure \ref{fig:nu-anti_nu_same_flavor}) and a $\delta_{\beta\beta'}$ contribution (Figure \ref{fig:nu-anti_nu_new_flavor}) , which is diagonal if $\beta=\alpha$ or off-diagonal if $\beta\neq\alpha$.

Again, considering a propagating neutrino $\nu_\alpha (p_1,s_1)$, we can write the potential as a matrix in the flavor space of this neutrino.

\begin{equation}
V_{\nu\overline{\nu}} = - \frac{ \sqrt{2} G_F}{V} (1-\hat{p}_1 \cdot \hat{p}_2)
\begin{pmatrix}
2\delta_{\alpha',e}+\delta_{\alpha',\mu}+\delta_{\alpha',\tau} & \delta_{\alpha',\mu} & \delta_{\alpha',\tau}\\ 
\delta_{\alpha',e} & \delta_{\alpha',e} +2\delta_{\alpha',\mu} + \delta_{\alpha',\tau} & \delta_{\alpha',\tau}\\ 
\delta_{\alpha',e} & \delta_{\alpha',\mu} & \delta_{\alpha',e} +\delta_{\alpha',\mu} + 2\delta_{\alpha',\tau}
\end{pmatrix}
\end{equation}

As we did in equation \ref{eq:Final_potential_nu_nu_with_diagonal}, we can remove the contribution proportional to the diagonal, finally getting the expression for the neutrino-antineutrino potential.

\begin{equation}
\label{eq:Final_potential_nu_antinu}
V'_{\nu\overline{\nu}} = - \frac{\sqrt{2}G_F}{V} (1-\hat{p}_1 \cdot \hat{p}_2)
\begin{pmatrix}
\delta_{\alpha',e} & \delta_{\alpha',\mu} & \delta_{\alpha',\tau}\\ 
\delta_{\alpha',e} & \delta_{\alpha',\mu} & \delta_{\alpha',\tau}\\ 
\delta_{\alpha',e} & \delta_{\alpha',\mu} &\delta_{\alpha',\tau}
\end{pmatrix}
\end{equation}

Note that this potential matrix in the flavor basis is, apart from a minus sign, the transpose from the neutrino-neutrino one in equation \ref{eq:Final_potential_nu_nu} when looking at the flavor content of the interacting neutrino $\nu_{\alpha'}$.

\section{Connection to Density Matrix Formalism}
\label{sec:Density_matrix}
Finally, let us connect the potentials derived above to the density matrix formalism. We will consider a neutrino quantum state $\ket{\psi_\nu (t)}$ going into a forward scattering with another neutrino in a state $\ket{\phi_\nu (t)}$ or antineutrino $\ket{\phi_{\overline{\nu}} (t)}$. Therefore, we can define the density matrix which represents the background of neutrinos as an incoherent mixture of each neutrino quantum state.
\begin{equation}
\label{eq:density_matrix_def}
    \rho_{p_2} (t) = \sum_{\nu'} n_{\nu',p_2} \ket{\phi_{\nu',p_2}(t)} \bra{\phi_{\nu',p_2}(t)} \rightarrow [\rho_{p_2} (t)]_{\alpha\beta} \equiv \bra{\nu_\alpha}\rho_{p_2} (t) \ket{\nu_\beta}
\end{equation}

\begin{equation}
\label{eq:anti-density_matrix_def}
    \overline{\rho}_{p_2} (t) = \sum_{\overline{\nu}'} n_{\overline{\nu}',p_2} \ket{\phi_{\overline{\nu}',p_2}(t)} \bra{\phi_{\overline{\nu}',p_2}(t)} \rightarrow [\overline{\rho}_{p_2} (t)]_{\beta\alpha}\equiv \bra{\overline{\nu}_\alpha}\overline{\rho}_{p_2} (t) \ket{\overline{\nu}_\beta}
\end{equation}

Where $\ket{\phi_{\nu',p_2}(t)}$ is the state of the neutrino with momentum $\vec{p}_2$ at time $t$ elapsed after its creation, and $n_{\nu',p_2}$ is density occupation number of this state. The index $\nu'$ represents any other quantum number for the neutrinos with momentum $\vec{p}_2$, such as flavor or mass eigenvalues. Note that we define the matrix elements $[\rho_{p_2} (t)]_{\alpha\beta} $ and $[\overline{\rho}_{p_2} (t)]_{\beta\alpha}$ with inverted indices, as adopted in \cite{sigl1993general,Duan:2010bg} . This allows $\rho$ and $\overline{\rho}$ to have the same transformation when changing from one basis to another. Note also that we could consider the density matrix of the entire neutrino ensemble by summing over the momentum $\vec{p}_2$. However, we will keep the momentum dependence, as it will be useful in the calculations below.

Considering the evolution equation \ref{eq:evolution} for the neutrino, we can work with each component of the momentum basis separately instead of a full wave packet as in equation \ref{eq:wave-packet}. We can do this simplification because we are considering only forward scattering. Therefore, the evolution equation for each eigenstate with momentum  $\vec{p}_1$ is the following in the flavor basis
\begin{equation}
\begin{split}
   i \frac{d}{dt} \sum_{\alpha} \psi_{\alpha}(t) \ket{\nu_{\alpha}(p_1)} =& \sum_{\alpha}  \psi_{\alpha}(t) H \ket{\nu_{\alpha}(p_1)} = \sum_{\alpha,\beta} \psi_{\alpha}(t) \ket{\nu_{\beta}(p_1)} \bra{\nu_{\beta}(p_1) }H \ket{\nu_{\alpha}(p_1)} \\
\end{split}
\end{equation}

To calculate the matrix element we have to consider the background final and initial states $\ket{\phi_{\nu',p_2} (t)}$ and then sum over the momentum $\vec{p}_2$ considering an occupation number distribution $N_{\nu',p_2}$. The same has to be done for the antineutrinos states $\ket{\phi_{\overline{\nu}',p_2} (t)}$. As we are interested in the neutrino-neutrino potential, we will ignore here the vacuum  and matter potential components of the Hamiltonian. 

\begin{equation}
\begin{split}
   &i \frac{d}{dt} \sum_{\alpha} \psi_{\alpha}(t) \ket{\nu_{\alpha}(p_1)} =\\&= \sum_{\nu' p_2} \sum_{\alpha,\beta} N_{\nu',p_2}\psi_{\beta}(t) \bra{\nu_{\beta}(p_1), \phi_{\nu',p_2}(t)}V_{\nu\nu} \ket{\phi_{\nu',p_2}(t),\nu_{\alpha}(p_1) } \ket{\nu_{\beta}(p_1)}\\
   \\&+ \sum_{\overline\nu' p_2} \sum_{\alpha,\beta} N_{\overline\nu',p_2}\psi_{\beta}(t) \bra{\nu_{\beta}(p_1), \phi_{\overline{\nu}_,p_2}(t)}V_{\nu\overline{\nu}} \ket{ \phi_{\overline{\nu}',p_2}(t), \nu_{\alpha}(p_1)} \ket{\nu_{\beta}(p_1)}\\
\end{split}
\end{equation}

Decomposing the background neutrino and antineutrino states into the flavor basis, we have

\begin{equation}
\begin{split}
     &i \frac{d}{dt} \sum_{\alpha} \psi_{\alpha}(t) \ket{\nu_{\alpha}(p_1)}=\\&=\sum_{\nu' p_2} \sum_{\substack{\alpha,\beta \\ \alpha',\beta'}} N_{\nu',p_2} \psi_{\alpha}(t) \phi^{*}_{\nu' \beta'}(t) \phi_{\nu' \alpha'}(t)\ket{\nu_{\beta}(p_1)} \bra{\nu_{\beta}(p_1), \nu_{\beta'}(p_2)}V_{\nu\nu} \ket{\nu_{\alpha}(p_1),\nu_{\alpha'}(p_2)}\\
     &+ \sum_{\overline{\nu}' p_2} \sum_{\substack{\alpha,\beta \\ \alpha',\beta'}} N_{\overline{\nu}',p_2} \psi_{\alpha}(t) \phi^{*}_{\overline{\nu}' \beta'}(t) \phi_{\overline{\nu}' \alpha'}(t) \ket{\nu_{\beta}(p_1)} \bra{\nu_{\beta}(p_1), \overline{\nu}_{\beta'}(p_2)}V_{\nu\overline{\nu}} \ket{\nu_{\alpha}(p_1),\overline{\nu}_{\alpha'}(p_2)}\\
\end{split}
\end{equation}
     
Using the results from sections \ref{sec:Nu-Nu Potential} and  \ref{sec:Neutrino-Antineutrino} for the Hamiltonian matrix element, we can write
     
\begin{equation}
\begin{split}
     &i \frac{d}{dt} \sum_{\alpha} \psi_{\alpha}(t) \ket{\nu_{\alpha}(p_1)}=\\&+ \sum_{\nu' p_2} \sum_{\alpha, \alpha'} \sqrt{2}G_F \frac{N_{\nu',\vec{p}_2}}{V} \psi_{\alpha}(t) \phi^{*}_{\nu'\alpha'}(t) \phi_{\nu'\alpha'}(t) (1-\hat{p}_1 \cdot \hat{p}_2) \ket{\nu_{\alpha}(p_1)}\\
    &+ \sum_{\nu' p_2} \sum_{\alpha, \alpha'} \sqrt{2}G_F \frac{N_{\nu',\vec{p}_2}}{V} \psi_{\alpha}(t) \phi^{*}_{\nu'\alpha}(t) \phi_{\nu'\alpha'}(t) (1-\hat{p}_1 \cdot \hat{p}_2) \ket{\nu_{\alpha'}(p_1)}\\
    & - \sum_{\overline{\nu}' p_2} \sum_{\alpha, \alpha'} \sqrt{2}G_F \frac{N_{\overline{\nu}',\vec{p}_2}}{V} \psi_{\alpha}(t) \phi^{*}_{\overline{\nu}'\alpha'}(t) \phi_{\overline{\nu}'\alpha'}(t) (1-\hat{p}_1 \cdot \hat{p}_2) \ket{\nu_{\alpha}(p_1)}\\
     & - \sum_{\overline{\nu}' p_2} \sum_{\alpha, \beta} \sqrt{2}G_F \frac{N_{\overline{\nu}',\vec{p}_2}}{V} \psi_{\alpha}(t) \phi^{*}_{\overline{\nu}'\beta}(t) \phi_{\overline{\nu}'\alpha}(t) (1-\hat{p}_1 \cdot \hat{p}_2) \ket{\nu_{\beta}(p_1)}\\
\end{split}
\end{equation}

Note that the terms proportional to $\ket{\nu_{\alpha}(p_1)}$ only contribute to the diagonal and with the same amplitude independent of the flavor $\alpha$. Therefore, these terms only contribute to an overall phase and can be ignored. Writing the result in the matrix format

\begin{equation}
\begin{split}
i\frac{d}{dt}
\begin{bmatrix}
\psi_{e }(t)\\ 
\psi_{\mu}(t)\\
\psi_{\tau}(t)
\end{bmatrix}
&= \sum_{\nu', \vec{p}_2}
\sqrt{2} G_F  (1-\hat{p}_1 \cdot \hat{p}_2) n_{\nu',\vec{p}_2}
\begin{bmatrix}
|\phi_{p_2}^{e}|^2  & \phi_{p_2}^{\mu *} \phi_{p_2}^{e} & \phi_{p_2}^{\tau *} \phi_{p_2}^{e}  \\
\phi_{p_2}^{e *} \phi_{p_2}^{\mu}& |\phi_{p_2}^\mu|^2 &\phi_{p_2}^{\tau *} \phi_{p_2}^{\mu}\\
\phi_{p_2}^{e *} \phi_{p_2}^{\tau}& \phi_{p_2}^{\mu *} \phi_{p_2}^{\tau} &|\phi_{p_2}^\tau|^2\\
\end{bmatrix}
\begin{bmatrix}
\psi_{e }(t)\\ 
\psi_{\mu}(t)\\
\psi_{\tau}(t)
\end{bmatrix}\\
&- \sum_{\overline{\nu}', \vec{p}_2}
\sqrt{2} G_F  (1-\hat{p}_1 \cdot \hat{p}_2) n_{\overline{\nu}',\vec{p}_2}
\begin{bmatrix}
|\phi_{p_2}^{e}|^2 & \phi_{p_2}^{e *} \phi_{p_2}^{\mu}  & \phi_{p_2}^{e *} \phi_{p_2}^{\tau} \\
\phi_{p_2}^{\mu *} \phi_{p_2}^{e} & |\phi_{p_2}^{\mu}|^2  &\phi_{p_2}^{\mu *} \phi_{p_2}^{\tau}\\
\phi_{p_2}^{\tau *} \phi_{p_2}^{e}& \phi_{p_2}^{\tau *} \phi_{p_2}^{e} &|\phi_{p_2}^{\tau}|^2 \\
\end{bmatrix}
\begin{bmatrix}
\psi_{e }(t)\\ 
\psi_{\mu}(t)\\
\psi_{\tau}(t)
\end{bmatrix}
\end{split}
\end{equation}

Where the occupation number $N_{\nu',\vec{p}_2}$ divided by the volume turns into a density occupation number $n_{\overline{\nu}',\vec{p}_2}$. Using the matrix density definition in equations \ref{eq:density_matrix_def} and \ref{eq:anti-density_matrix_def} , we can rewrite this as

\begin{equation}
\label{eq:Final_potential_density_matrix}
  i\frac{d}{dt} \ket{\psi_{\nu}(t)}  =\sum_{\vec{p}_2}
\sqrt{2} G_F (1-\hat{p}_1 \cdot \hat{p}_2) [\rho_{\vec{p}_2}(t)-\overline{\rho}_{\vec{p}_2}(t)] \ket{\psi_{\nu}(t)}
\end{equation}
This is finally the neutrino-neutrino potential format usually used in other works. There are alternative definitions of the density matrix that may lead to a slightly different expression, depending on other variables rather than the momentum. However, this momentum-dependent density matrix can easily be translated into the neutrino spectrum and vice-versa, which is helpful for many applications.

\subsection{Polarization Vector Formalism}

One useful application for the neutrino-neutrino potential in the format of equation \ref{eq:Final_potential_density_matrix} is the so-called Polarization Vector formalism. If we consider only two families of neutrinos (e.g. $\nu_e$ and $\nu_x$ \footnote{This can be applied to supernovae, given that $\nu_\mu$ and $\nu_\tau$ are produced in the same quantity and have the same interactions in the propagation and detection.}) the matrices in the evolution equation will be $2\times 2$ complex matrices, which can be decomposed in Pauli matrices $(\sigma_1,\sigma_2, \sigma_3)$. The coefficients of this expansion work as components of a vector in a 3-dimensional space, which are called Polarization Vectors. To see this, let us rewrite the evolution of the neutrino state $\ket{\psi_{\nu}(t)}$ in terms of its density matrix.

\begin{equation}
    i\frac{d\rho (t)}{dt} = [H , \rho (t)], \;\;\; \rho(t) \equiv \ket{\psi_{\nu}(t)}\bra{\psi_{\nu}(t)}
\end{equation}

We can therefore decompose the system in Pauli matrices.

\begin{equation}
     H = -\frac{1}{2}\vec{\sigma} \cdot \vec{B} \text{,} \;\;\;\;\; \rho= \frac{1}{2} \mathbf{1} + \frac{1}{2} \vec{\sigma} \cdot \vec{P}\text{,} \;\;\;\;\; \vec{\sigma}=(\sigma_1,\sigma_2,\sigma_3)
\end{equation}

Note that we ignore the Hamiltonian component proportional to the identity, which is irrelevant for the evolution. With this decomposition, the evolution equation for the neutrino became a precession equation due to the anti-commutation rules of the Pauli matrices. In this picture, the polarization vector of the neutrino $\vec{P}$ precesses around the Hamiltonian vector $\vec{B}$.

\begin{equation}
      \frac{d}{dt} \vec{P}(t)= \vec{P}(t)\times \vec{B},  \;\;\;\;\;  \left ( i\frac{d}{dt} P_i(t) = \frac{i}{2} P_i(t) B_j [\sigma_i,\sigma_j] \right)
\end{equation}

In the case of neutrino-neutrino interactions, considering $H=V_{\nu\nu} + V_{\nu \overline{\nu}}$, we have the following evolution equation.

\begin{equation}
    \frac{d}{dt}\vec{P}_{\nu,\vec{p_1}} = \vec{P}_{\nu,\vec{p_1}} \times \sum_{\vec{p}_2}  \sqrt{2} G_F (1-\hat{p}_1 \cdot \hat{p}_2) (\vec{P}_{\nu,\vec{p}_2} - \vec{P}_{\overline{\nu},\vec{p}_2})
\end{equation}

Therefore, a given neutrino polarization vector precesses around all the others. This gives us a good pictorial view of the neutrino evolution inside a high-density neutrino environment. Besides, this formalism is highly used in works trying to solve the neutrino evolution in a high-density neutrino environment, such as in supernovae. 

\section{Conclusions}
\label{sec:Conclusions}

In this paper, we have derived the potential due to neutrino-(anti)neutrino forward scattering at the low energy regime. In our approach, we took the Standard Model Hamiltonian for neutral current interactions in terms of the neutrinos second quantized field and found the matrix element of it given the initial and final states. The final expressions can be found in equations \ref{eq:Final_potential_nu_nu} for neutrino-neutrino interactions and \ref{eq:Final_potential_nu_antinu} for neutrino-antineutrino, or in equation \ref{eq:Final_potential_density_matrix} using the density matrix formalism.

To the best of the authors' knowledge, this is the first time that these calculations were explicitly done using massive neutrino fields at the interaction. This approach shows explicitly how the mass of the neutrino appears in the potential expression. As expected, the neutrino mass contribution is suppressed by the energy in the ultra-relativistic limit, given the well-known result of the massless approximation.

The expression for the potential found in this paper can be applied for neutrinos in the energy regime of $E>>m_\nu$ (ultra-relativistic) and $E<m_{Z_0}$ (low energy). That is the case for supernova neutrinos, which was our motivation to derive this potential. The authors intend to use these results in future works searching for solutions to collective oscillations in supernova neutrinos. With the potential in terms of the density matrix, we can easily translate it to the neutrino polarization vectors formalism, in which we intend to explore the neutrino evolution in a supernova environment.

\appendix

\section{Conventions}
\label{sec:Convetions}
Here we describe our conventions regarding the neutrinos quantized fields and states. We take the following Fourier expansion for the quantized neutrino fields, using a finite normalization volume $V$ with periodic conditions,

\begin{equation}
\label{eq:field_expansion_V}
\begin{split}
        \nu_\alpha(x)&= \sum_{\vec{p},s}  \frac{1}{\sqrt{2VE_{\vec{p}}}} \left [ a^s_{\vec{p}} u^s(\vec{p}) e^{-ipx} +  b^{s \dagger}_{\vec{p}} v^s(\vec{p}) e^{ipx}\right ]
\end{split}
\end{equation}

\begin{equation}
\label{eq:field_expansion_adj_V}
\begin{split}
        \overline{\nu}_\alpha(x)&= \sum_{\vec{p}, s}  \frac{1}{\sqrt{2VE_{\vec{p}}}} \left [ a^{s \dagger}_{\vec{p}} \overline{u}^s(\vec{p}) e^{ipx} +  b^{s}_{\vec{p}} \overline{v}^s(\vec{p}) e^{-ipx}\right ]
\end{split}
\end{equation}

where the anti-commutation relations are
\begin{equation}
    \{  a^{s \dagger}_{p},  a^{s' \dagger}_{p'} \} = \{  b^{s \dagger}_{p},  b^{s' \dagger}_{p'} \} =\delta_{\vec{p}\vec{p}'} \delta_{ss'}
\end{equation}

Therefore, the fields obey the canonical equal-time anticommutation rules

\begin{equation}
    \{ \nu_\alpha (t,\vec{x}), \nu^{\dagger}_\beta (t,\vec{y}) \} = \delta^{(3)} (\vec{x} - \vec{y}) \delta_{\alpha \beta}
\end{equation}

The one particle states are defined as
\begin{equation}
    \ket{\nu(p,s)}=  a^{s \dagger}_{\vec{p}} \ket{0}   \text{,} \;\;\;\;\;  \ket{\overline{\nu}(p,s)}=  b^{s \dagger}_{\vec{p}} \ket{0}
\end{equation}

where their normalization is given by
\begin{equation}
    \bra{\nu(p,s)}\ket{\nu(p',s')} =  \bra{\overline{\nu}(p,s)}\ket{\overline{\nu}(p',s')} = \delta_{\vec{p}\vec{p}'} \delta_{ss'}
\end{equation}

Note that for Majorana neutrinos $a^{s \dagger}_{p}=b^{s \dagger}_{p}$.

\section{Trace Techniques}
\label{sec:trace_tec}
Here we derived some properties of spinors and Dirac matrices that are useful to our work. The resulting properties may be found by the name of "trace techniques" in the literature, as they translate the product of spinors and Dirac matrices into trace expressions.

\subsection{Spinor to Trace}

Lets consider a spinor that obeys the Dirac equation

\begin{equation}
    (\not{p} - m) u^{(h)}(p) = 0
\end{equation}
\begin{equation}
    \overline{u}^{(h)}(p) (\not{p} - m) = 0
\end{equation}

The helicity projector can be written as

\begin{equation}
    P_{h} = \frac{1+\gamma^{5}\not{s}_{h}}{2}
\end{equation}

where

\begin{equation}
    s^{\mu}_{h} = h \left( \frac{|\vec{p}|}{m} , \frac{E}{m} \frac{\vec{p}}{|\vec{p}|}\right)
\end{equation}

 Also, lets consider the completeness relation

\begin{equation}
\label{eq:spinor_completeness}
    \sum_{h=1}^{2} u^{h}(p) \overline{u}^{h}(p) = P_{h=1} u^{h}(p) \overline{u}^{h}(p) + P_{h=2} u^{h}(p) \overline{u}^{h}(p) = \not{p}+m
\end{equation}
Note that this expression is a $4\times4$ matrix, which can be seen due to the Dirac matrix in $\not{p}=\gamma^{\mu}p_{\mu}$. In general, all the vertices factors in SM interactions can be written as $ \overline{u}^{h}(p) \Gamma  u^{h}(p)$ where $\Gamma$ is a combination of Dirac matrices. We can use the relations above to turn this interaction factor into a trace expression.

\begin{equation}
\begin{split}
\label{eq:trace_projection}
    \overline{u}^{(h)}(p) \Gamma  u^{(h)}(p) =& \sum_{m,n}  \overline{u}^{(h)}_{n}(p) \Gamma_{nm}  u^{(h)}_{m}(p) = \sum_{m,n}   u^{(h)}_{m}(p) \overline{u}^{(h)}_{n}(p) \Gamma_{nm} \\
    =& \sum_{m,n}  \left[ P_{h} (\not{p}+m) \right]_{mn} \Gamma_{nm} = \sum_{m}\left[ P_{h}  (\not{p}+m) \Gamma \right]_{mm} \\
    =& Tr\left[ P_{h} (\not{p}+m) \Gamma \right] = Tr\left[ \frac{1+\gamma^{5}\not{s}_{h}}{2} (\not{p}+m) \Gamma \right]\\
    =& Tr\left[ (\not{p}+m) \left(\frac{1+\gamma^{5}\not{s}_{h}}{2} \right) \Gamma \right] 
\end{split}
\end{equation}
 
 Note that going from the first line to the second, we have used only one helicity component of equation \ref{eq:spinor_completeness}. Note also that $[\gamma^{5}\not{s}_{h},\not{p}]=0$. In the same way, we can derive the following expression for the sum over the helicity states.
 
 \begin{equation}
 \label{eq:trace_sum}
    \sum_{h=1}^{2} \overline{u}^{(h)}(p) \Gamma  u^{(h)}(p) = Tr\left[ (\not{p}+m) \Gamma \right] 
 \end{equation}
 
 \subsection{Weak SM interactions}
In the specific case of weak interactions of the Standard Model, we have the following V-A structure $\Gamma= \gamma^{\mu}(1-\gamma^5)$. Therefore, equation \ref{eq:trace_projection} became 
 
\begin{equation}
\begin{split}
 \overline{u}^{(h)}(p) \gamma^{\mu}(1-\gamma^5)  u^{(h)}(p) =&  \frac{1}{2} Tr\left[(\not{p}+m) (1+\gamma^{5}\not{s}_{h}) \gamma^{\mu}(1-\gamma^5) \right]=\\
 =& \frac{1}{2} Tr[(\not{p}+m)(\gamma^{\mu}-\gamma^{\mu}\gamma^{5}+\gamma^{5}\not{s}_{h}\gamma^{\mu}-\gamma^{5}\not{s}_{h}\gamma^{\mu}\gamma^{5})]=\\
 =& \frac{p_{\nu}}{2} Tr[\gamma^{\nu}\gamma^{\mu}-\gamma^{\nu}\gamma^{\mu}\gamma^{5}+\gamma^{\nu}\gamma^{5}\gamma^{\sigma} s^{h}_{\sigma}\gamma^{\mu}-\gamma^{\nu}\gamma^{5}\gamma^{\sigma} s^{h}_{\sigma}\gamma^{\mu}\gamma^{5}]\\
  +& \frac{m}{2} Tr[\gamma^{\mu}-\gamma^{\mu}\gamma^{5}+\gamma^{5}\gamma^{\sigma} s^{h}_{\sigma}\gamma^{\mu}-\gamma^{5}\gamma^{\sigma} s^{h}_{\sigma}\gamma^{\mu}\gamma^{5}]
\end{split}
\end{equation}

Using the following proprieties\footnote{And the trace property $ Tr[A+B] = Tr[A] + Tr[B]$.}

\begin{equation}
\begin{split}
    &Tr[\gamma^{\nu}\gamma^{\mu}] = 4 g^{\mu\nu}\\
    &Tr[\gamma^{\nu}\gamma^{\mu}\gamma^{5}] = 0\\
    &Tr[\gamma^{\nu}\gamma^{5}\gamma^{\sigma}\gamma^{\mu}] = 0\\
    &Tr[\gamma^{\nu}\gamma^{5}\gamma^{\sigma}\gamma^{\mu}\gamma^{5}] = Tr[\gamma^{\nu}\gamma^{\sigma}\gamma^{\mu}]= 0\\
\end{split}
\end{equation}

\begin{equation}
\begin{split}
    &Tr[\gamma^{\mu}] = 0\\
    &Tr[\gamma^{\mu}\gamma^{5}] = 0\\
    &Tr[\gamma^{5}\gamma^{\sigma}\gamma^{5}] = 0\\
    &Tr[\gamma^{5}\gamma^{\sigma}\gamma^{\mu}\gamma^{5}] = Tr[\gamma^{\sigma}\gamma^{\mu}]= 4g^{\sigma\mu}\\
\end{split}
\end{equation}

we finally have our desired relation.

\begin{equation}
\label{eq:A_spinor_projection}
    \overline{u}^{(h)}(p) \gamma^{\mu}(1-\gamma^5)  u^{(h)}(p) = 2(p^{\mu} - m s^{\mu}_h)
\end{equation}

We can use the same reasoning for the equation \ref{eq:trace_sum} and obtain the relation bellow.

\begin{equation}
\label{eq:A_spinor_sum}
    \sum_{h=1}^2 \overline{u}^{(h)}(p) \gamma^{\mu}(1-\gamma^5)  u^{(h)}(p) = 4p^{\mu}
\end{equation}

\section{About the Potential Normalization}
\label{sec:Potential Normalization}
The one-particle states of momentum $\vec{p}$ are defined in a way that their normalization is Lorentz invariant. 

\begin{equation}
\label{eq:B_normaliztion}
    \bra{\vec{q}}\ket{\vec{p}} = (2\pi)^3 2E_{\vec{p}} \delta^{(3)} (\vec{p}-\vec{q})
\end{equation}

However, as in quantum mechanics, these states are not properly normalized to one in the continuum regime, and therefore can not describe physical particles. One way to get rid of this is to use wave packets properly normalized to represent particles. Another way is to consider finite volume normalization with periodic boundary conditions. That shows us that amplitudes calculated using states such as in \ref{eq:B_normaliztion} will not be properly normalized. However, the right normalization using these states can be achieved by the following process.

\begin{equation}
    \mathcal{A}_{norm} = \frac{\mathcal{A}_{fi}}{\sqrt{\bra{f}\ket{f}\bra{i}\ket{i}}}= \frac{\mathcal{A}_{fi}}{\prod_{i}^{} \sqrt{E_i} \prod_{f}^{} \sqrt{E_f}} = \frac{\bra{f}\mathcal{S}\ket{i}}{\sqrt{\bra{f}\ket{f}\bra{i}\ket{i}}}
\end{equation}

\section*{Acknowledgments}

This work was supported by São Paulo Research Foundation (FAPESP) grants no. 2019/08956-2 and no. 14/19164-6.

\bibliographystyle{ws-mpla}
\bibliography{sample}
\end{document}